\documentclass[%
 reprint,
 amsmath,amssymb,
 aps,
pre,
]{revtex4-1}
\pdfoutput=1

\usepackage{dcolumn}
\usepackage{bm}
\usepackage{hyperref}

\usepackage{subfigure}
\usepackage[pdftex]{graphicx}

\hypersetup{ colorlinks,
linkcolor=blue,
filecolor=blue,
urlcolor=blue,
citecolor=blue }

\begin{document}


\title{Epidemic SIR model on a face-to-face interaction network: new mobility induced phase transitions}

\author{P. F. Gomes}
\email{paulofisicajatai@gmail.com}
 \altaffiliation{Present address: Instituto de F\'isica de S\~ao Carlos, Universidade de S\~ao Paulo, S\~ao Carlos-SP, Brazil.}
\author{H. A. Fernandes}
\author{A. G. Fran\c{c}a}
\author{F. L P. Costa}
\affiliation{%
Instituto de Ci\^encias Exatas, Universidade Federal de Jata\'i, BR 364, km 195, 75801-615, Jata\'i-GO, Brazil.
}%


\author{R. S. Grisotto}
\affiliation{Instituto de Computa\c{c}\~ao, Universidade Estadual de Campinas, Campinas-SP, Brazil.
}%

\date{\today}

\begin{abstract}
In this work, we study the epidemic SIR model on a system which takes into consideration face-to-face interaction networks. This approach has been used as prototype to describe people interactions in different kinds of social organizations and, here, it is considered by means of three features of human interactions: the mobility, the duration of the interaction among people, and the dependence of the number of interactions of each person on the time evolution of the system. For this purpose, the initial configuration of the system is set as a regular square lattice where the nodes are the individuals which, in turn, are able to move in a random walk along the network. So, the connectivity among the individuals evolve with time and is defined by the positions of the individuals at each iteration. In a time unit, each individual is able move up to a distance $v$ creating different networks along the time evolution of the system. In addition, the individuals are interacting with each other only if they are within the interaction distance $\delta$ and, in this case, they are considered as neighbors. If a given individual is interacting with other ones, he performs the random walk with a diffusion probability $\omega$. Otherwise, the diffusion occurs with probability 1. The study was carried out through nonequilibrium Monte Carlo Simulations and we take into account the asynchronous updating scheme. The results show that, for a given $v>0$, there exist a critical line in the $(c, \delta)$ space, where $c$ is the immunization rate. We also obtain the dynamic critical exponent $\theta$ for some points belonging to this line and show that this model does not belong to the directed percolation universality class.
\end{abstract}

\pacs{Valid PACS appear here}
\maketitle


\section{Introduction}

The dynamics of the time evolution in a given population is a determinant aspect for the outcome of processes such as gossip/opinion dissemination and scientific-collaboration interactions \cite{Fortunato2018}. Different models for social networks has been implemented in order to understand the intrincated mechanism of human interactions, for instance, scale free \cite{Barabase1999} and small world \cite{Watts1998} networks. However the frequency and time duration of the interactions between two people in a human gathering (conferences, schools, or museums for example) play a crucial role in the collective behaviour of the group. The so called face-to-face networks have been used in order to mimic these important features in the study of processes such as online communication and epidemic spreading \cite{Wang2014,Pozzana2017,Starnini2013}.

Epidemic models have played an important role in statistical mechanics due to their rich dynamics \cite{Hethcote2000,PastorSatorras2001,Mieghem2013}, as well as their importance in complex systems \cite{Earn2000,Vespignani2012} and networks \cite{Newman2002,Keeling2005}. Additionally, experimental data (gathered from conferences and places such as schools or hospitals) have been used to model the spreading of diseases \cite{Genois2015,Fournet2016}, openning new possibilities for studying the relevant details of their intricate mechanisms. Out of equilibrium, the time behavior of the epidemic outbreak has been studied with short-time Monte Carlo simulations \cite{Silva2015} and also within the context of complex networks \cite{Barthelemy2004,Barthelemy2005}.

For this model, it has been shown that the finite size scaling near criticality follows the general scaling relation given by \cite{Hinrichsen2000}:
\begin{equation}
\langle i(t) \rangle \sim t^{-\beta / \nu_{\parallel}} f\left(  \Delta t^{1/v_{\parallel}}, L^{-d} t^{d/z}, d_0 t^{\theta + \beta / v_{\parallel}}  \right), \nonumber
\end{equation}
where $i(t)=I(t)/N$ is the density of infected individuals in a $d-$dimensional lattice at the time $t$ ($I(t)$ is the total number of individuals), $N$ is the number of individuals placed in a square box of linear size $L$, $\Delta=c-c_c$ measures the distance from a point $c$ (recovery rate) to the critical point (critical recovery rate), and $\langle \cdot \cdot \cdot \rangle$ means the usual average on different $N_s$ samples of the system. The exponents $\beta$, $\nu_{\parallel}$, and $\nu_{\perp}$ are static critical exponents, while $z = \nu_{\parallel}/\nu_{\perp}$ and $\theta=\frac{d}{z}-\frac{2\beta}{\nu_{\parallel}}$ are dynamic ones. One of the epidemic models which belongs to the directed percolation (DP) universality class is the susceptible-infected-susceptible (SIS) model that represents diseases for which the infection does not confer immunity, i.e., the individual returns to the susceptible class after being infected with the disease. In this case, at criticality $\Delta = 0$, it is expected two different power-law behaviors:
\begin{equation}
\langle i(t) \rangle = \left\{
\begin{array}{rl}
 t^{\theta}, \text{ if} \quad t < t_0   \\
 \\
t^{-\beta / z\nu_{\parallel}}, \text{ it} \quad t>t_0
\end{array} 
 \right.
 \label{Eq:powerlaws}
\end{equation}
The regime $t<t_0$ is obtained considering only one infected individual in the initial state, $t=0$, i.e., $i(0)=1/N$ and all other sites are occupied by susceptible ones. The second behavior, $t>t_0$, can be obtained by making $i(0)=1$ which means a network completely filled with infected agents.

In addition to the SIS model, there is the SI model (with no recovery), which is suitable for the study, for instance, of HIV virus. If the individual recovers of the disease and becomes immune to reinfection, one may consider the SIR model \cite{Grassberger1983}. However, if the recovering does not confer immunity, a reinfection can occur after some time of the recovery and this situation is considered in the SIRS model \cite{Souza2010}. In addition, there are other models which involve other states such as the exposed (E) state (for instance the SER \cite{Wada2015,Alexander2015} and SEIR \cite{Earn2000} models), as well as hidden (H) and maternally-derived immunity states \cite{Hethcote2000}.

In this work, we are concerned with the SIR model which is a well known model in the theory of epidemics \cite{Newman2002,Colizza2007,Argolo2011,Silva2015,Shu2015, Fournet2016,Estrada2016} in part due to its interesting phase diagram and because it is a good model to mimic infectious diseases like measles, mumps, and rubella. In this model, each individual can be in one of three states: susceptible ($S$), infected ($I$), and recovered ($R$) and there are two distinct phases: The first phase is obtained when the recovery (immunization) rate is large enough to prevent the epidemic spreading throughout the lattice (in this case there is only the formation of small clusters of recovery individuals); the second phase occurs when the immunization rate is small and, therefore, there is an outbreak of the disease. In this case, a large component of the network becomes infected ($i(t)$ increases up to a maximum value) and afterwards recovered ($i(t)$ decreases down to zero). The point which separates the non-spreading of the spreading of the epidemic is known as the phase transition point of the model. In this point, we expect the power law $i(t) \sim i_0 t^{\theta}$ to be satisfied. 

The paper is organized as follows: in the next section we describe the standard SIR model as well as our extended version of the model considering the modified connectivity (dependent on $v$ and $\delta$). In Sec. \ref{sec:results}, we present the numerical simulation technique used in this work along with our main numerical results. Finally, the summaries and conclusions are considered in Sec. \ref{sec:conclusions}. The appendixes present further information about the coefficient of determination approach and the computational implementation of the temporal evolution of the system.

\section{Epidemic SIR model on a face-to-face network}
\label{sec:model}

The traditional set of differential equations for the epidemic SIR model is given by \cite{Newman2010}:
\begin{eqnarray}
\dfrac{dS(t)}{dt} = - \beta IS, \qquad \dfrac{dI(t)}{dt} = \beta IS - \gamma I, \qquad \dfrac{dR}{dt} &=&  \gamma I, \nonumber 
\end{eqnarray}
where $S(t)$, $I(t)$, and $R(t)$ are the number of susceptible, infected, and recovered individuals, respectively, as function of the time $t$, $\beta$ is the contact rate which is related to the probability of a susceptible individual being infected with the disease, and $\gamma$ is the recovery rate of infected individuals. From these equations, one can see that $\frac{d}{dt}[S(t)+I(t)+R(t)]=\frac{dN(t)}{dt}=0$, i.e., the number of individuals in the lattice, $N(t)$, is constant at any time. In our simulations, we consider that the rates $\beta$ and $\gamma$ are given, respectively, by the infection $b$ and the recovery $c$ probabilities. We also consider that $b+c=1$ \cite{Arashiro2007} and $s = S/N$, $i = I/N$, and $r = r/N$ are the density of susceptible, infected, and recovered individuals, respectively.

This approach supposes a full connected network, i.e., each agent uniformly interacts with all the other agents (homogeneous mixing hypothesis \cite{Barthelemy2005}). However this connectivity can be altered by different criteria, for instance, the distance between the agents considering the system domain as a square in the 2D Euclidean plane $xy$. This can be done in two ways: i) by treating the interaction probability as a continuous variable that decays with increasing distance \cite{Juhasz2015} or ii) by considering two individuals interacting when the distance between them is less than a specific value $\delta$. In a regular network (as, for instance, a regular square lattice) one can consider only the interactions among an individual and his neighbors (nearest ones, next-nearest ones, etc.) \cite{Silva2015}. In a heterogeneous connectivity pattern network, the interactions can be built in any desired way, placing the individuals on the vertices and representing the interaction with an edge \cite{Newman2010}.


Many works have explored the effect of the mobility of the individuals on epidemic models since it has been suggested in 1983 \cite{Grassberger1983}. For example, the SIR model has been studied by considering both the diffusion between different subpopulations \cite{Colizza2007} (a population in only one node) and the diffusion between nearest-neighbors in a diluted regular square lattice \cite{Silva2015}. There are still studies which consider the mobility between specific places (or nodes) emulating house and work \cite{Belik2011} and the mobility among communities (group of nodes) forming a general network with real human commuting data \cite{Balcan2011,Ren2014}. However, none of them have allowed the individuals to move freely in the network, forming new random temporal networks at each iteration. Here, one iteraction means one Monte Carlo step, i.e., $N$ updates (trials) of the states/positions of the $N$ individuals of the network. One such way to accomplish this movement is to allow each vertex (or individual) to perform a random walk: in each iteration, each node can move a distance up to $v$ in an arbitrary direction $[0,2\pi]$. Indeed, random walk has been frequently used in temporal networks \cite{Alessandretti2017,Starnini2013,Scholtes2014,Perra2012} as well as in critical phenomena \cite{Dorogovtsev2008}. It is worth to mention that, as the interaction between two vertices occurs only when their distance is smaller than a specific value $\delta$, each new interaction changes the interacting pairs.


Every time-dependent Monte Carlo (MC) simulation starts with a regular square lattice where all individuals are susceptible to the disease but one which is already infected \cite{Silva2015,Souza2010,Tome2010}. In one iteration, each individual can perform a random walk with the displacement up to $v$ in an arbitrary direction $[0,2\pi]$. The update of both state and position of each individual is performed by using the asynchronous updating algorithm. When the Euclidean distance $d$ between two individuals is smaller than or equal to $\delta$, they are interacting and considered as neighbors. In this case, they may continue in their positions (not walking) with a probability $1-\omega$, where $\omega$ is the diffusion probability when the individuals are interacting. However, if $d>\delta$, the two individuals are not interacting with each other and, therefore, they will move with probability 1. One susceptible individual can become infected if he interacts with at least one infected individual. The probability of infection is $bn_i/n_t$ where $n_i$ is the number of infected neighbors and $n_t$ is the total number of neighbors. Both $n_i$ and $n_t$ change for each individual and with time $t$. On the other hand, as usually happens in the original model, an infected individual recovers from the disease with a rate $c$. The other important parameters of the model are: $\rho = N/L^2$, the density of individuals; $Q$, the number of Monte Carlo steps; and $N_S$, the number of independent samples of each simulation. We also consider periodic boundary conditions in our simulations. To retrieve the standard SIR model, in which the individuals are not able to move, we must simply set $v=0$. In this case, it has been found that the critical immunization rate is $c_0 = 0.1765$ \cite{Souza2010,Tome2010}.

\section{Numerical simulations and Results}
\label{sec:results}

As we are studying an extended version of the SIR model, we first wonder if its critical behavior changes with the modified connectivity, thus changing its universality class. To answer our question, we perform time-dependent Monte Carlo simulations along with an optimization method which takes into account a concept known as coefficient of determination defined as \cite{Trivedi2002}:
\begin{equation}
\alpha = \dfrac{  \sum_{t=t_c}^Q \left( p+q \ln t - \ell \right)^2 }{ \sum_{t=t_c}^Q \left( \ln \langle i(t) \rangle - \ell \right)^2  }. \label{coefdeterminacal}
\end{equation}
where
\begin{equation*}
\ell = \dfrac{1}{Q-t_c+1} \sum_{t=t_c}^Q \ln i(t)
\end{equation*}
and
\begin{equation*}
\langle i(t) \rangle = \dfrac{1}{N_S} \sum_j^{N_S} i_j(t)
\end{equation*}
are, respectively, the average number of infected individuals in the time range and the average of $i(t)$ for different samples (more details are found in Appendix \ref{ap:a}). The coefficient is calculated for $t>t_c$.  

This coefficient measures the ratio between the expected variation and the total variation. It takes into consideration the robustness of the power laws defined at criticality since $\alpha \simeq 1$ when the system is close to its critical point (and $i(t) \sim t^{\theta}$) and $\simeq 0$ otherwise. In summary, if the data follows a linear behavior (in log scale), the system is probably at its critical point and the expected variation is close to the total variation, and $\alpha \simeq 1$. This makes easier to find the phase transition point: just look for points where $\alpha \simeq 1$. So, by using this optimization method, we are able to identify the set of values for $c$ and $\delta$ where the phase transition of the system occurs, for a given mobility $v$. This approach was first proposed in nonequilibrium MC simulations by da Silva \textit{et al.} \cite{Silva2012} in the context of generalized statistics. After that work, the phase transitions of several models have been successfully identified, such as models with defined Hamiltonian \cite{Silva2013,Silva2013b,Fernandes2014,Fernandes2017} and models with absorbing states \cite{Silva2015,Fernandes2016}. 

The density of individuals used in all simulations presented in this work is $\rho = 512 \times 10^{-4}$ (equivalent to 512 individuals in a square with linear size 100) and the lattice parameter of the initial square network is $a=L/\sqrt{N}$. In addition, we define the distance of interaction $\delta$ as a function of $a$. The density of individuals are calculated from the average of $N_s$ independent samples. Further information about the temporal evolution of the system is given in Appendix \ref{ap:b}.

Figures \ref{rede_0} and \ref{rede_23} show, respectively, the network for $t=0$ and $t=23$, for $N = 16^2$ and $\delta = 11a/10$. 
\begin{figure}
\centering
\subfigure{(a)
\includegraphics[width=2.6 in]{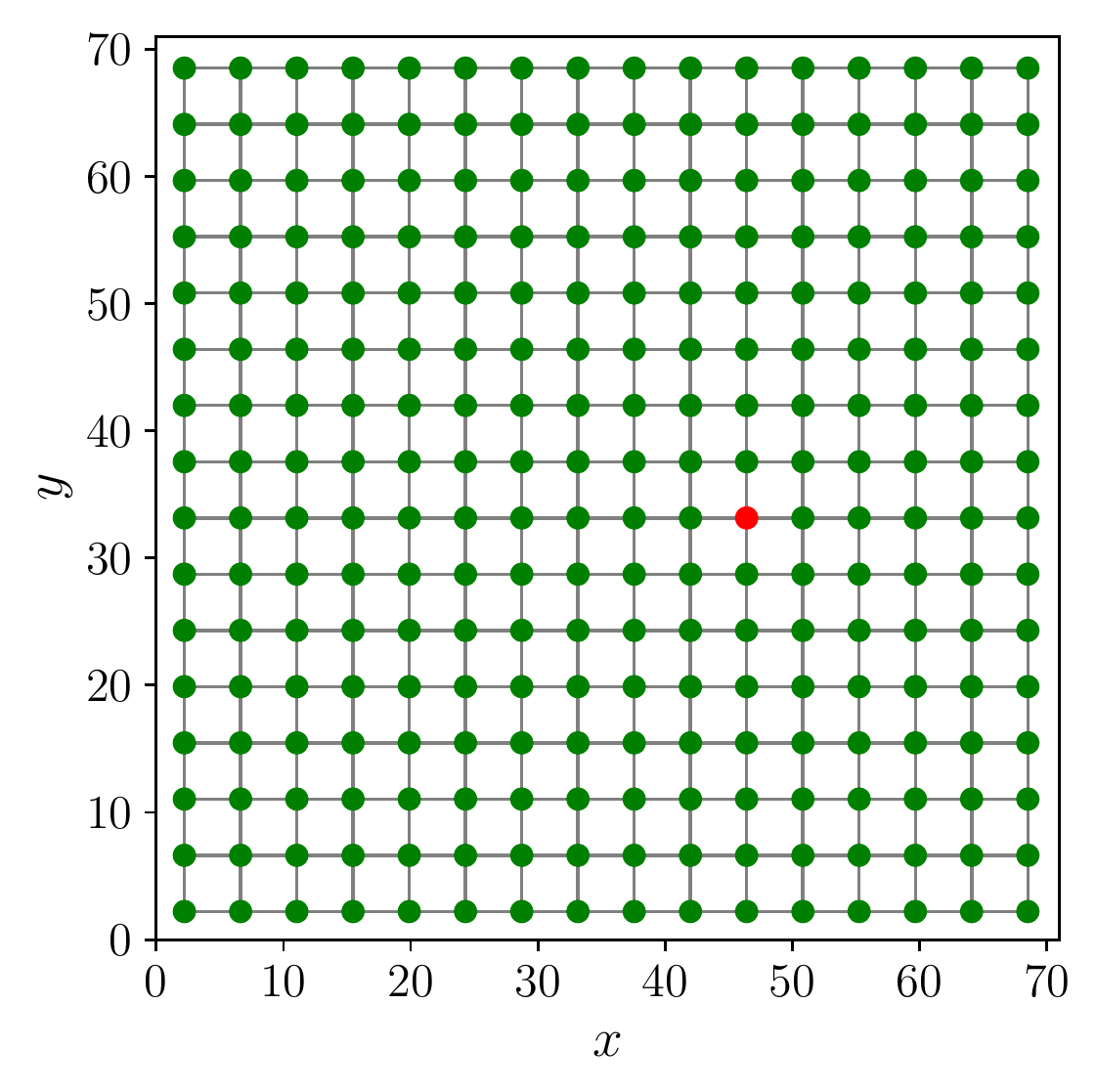}
\label{rede_0} 
}
\subfigure{(b)
\includegraphics[width=2.6 in]{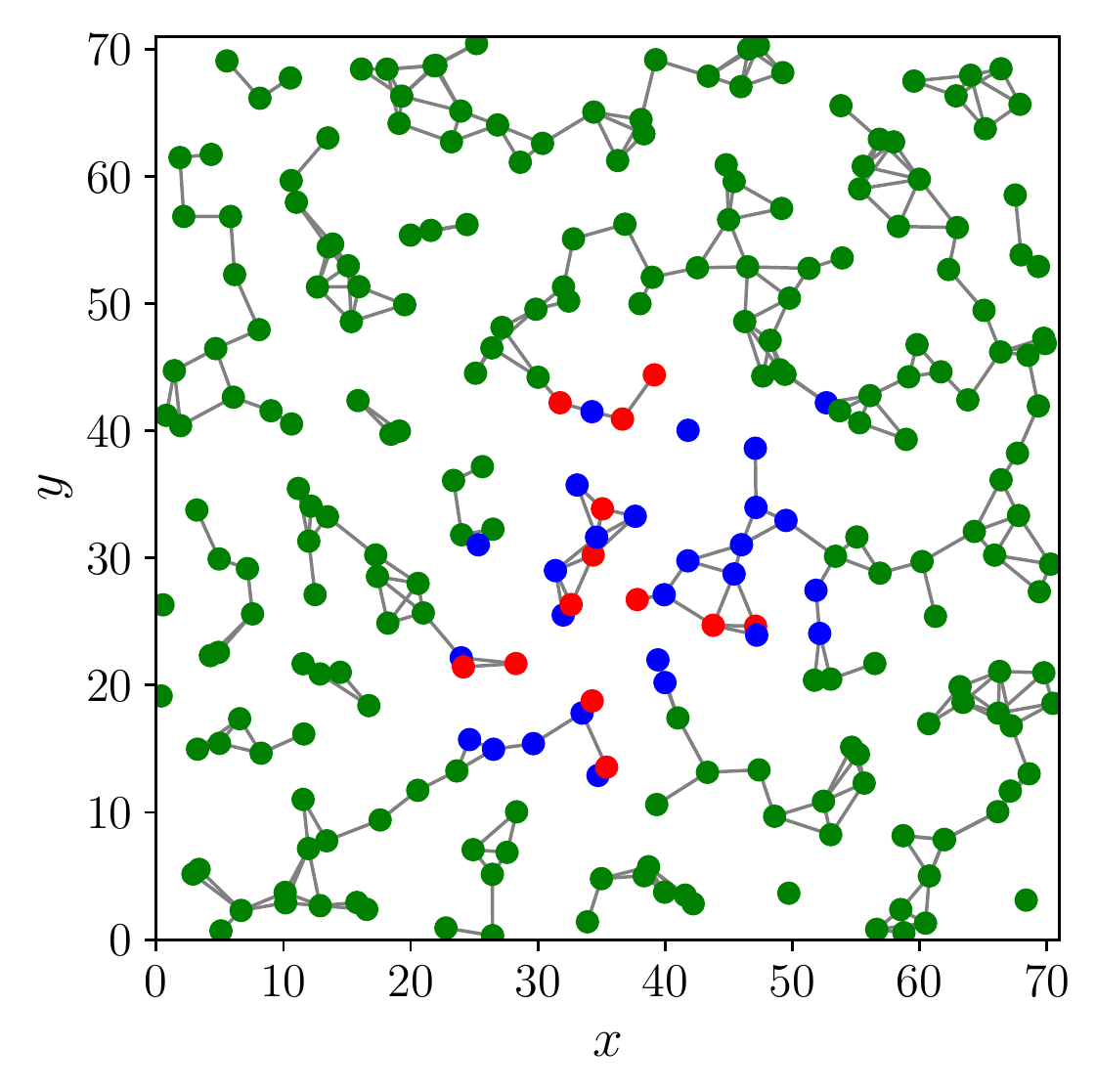}
\label{rede_23}
}
\caption{Snapshots of the network with $N=256$ individuals and $L=\sqrt{N/\rho} \approx 70.7$. The parameters are $d = 11a/10$ (so that only first neighbors interactions are considered). Figure \subref{rede_0}: Ordered grid at $t=0$. Figure \subref{rede_23}: Network at $t=23$ when the positions are changed. The dots with the green, red, and blue colors represent the susceptible, infected, and recovered individuals, respectively. The interactions are represented by the lines connecting the individuals.}
\end{figure}
As can be seen, at the very beginning, each individual interacts with the four nearest neighbors and, after some time, the random walk allows the individuals interact with different numbers of neighbors.

From now on, all the simulations are performed with the following set of parameters: $N=128^2$, $N_s=100$, and $\omega = 0.5$, which determine the linear size of the network, $L = \sqrt{N/\rho}\simeq 565.68$, and the lattice parameter $a=L/\sqrt{N} \simeq 4.42$.

The numerical simulations are performed by considering an initial condition in which all agents are in the susceptible state but one agent, chosen at random, that is infected with the disease, i.e.,  $s(0)=(N-1)/N$ and $i(0)=1/N$. In addition, the individuals can move through the network, and the way we include the mobility is allowing them to perform a random walk with a displacement $v$ in an arbitrary direction. When $v>0$ the square regular lattice is destroyed (for $t>0$) and a new temporary network is formed in each iteration. The Fig. \ref{rede_23} shows the system for $t=23$ with a displacement $v=1.0$ so that the number of interacting neighbors of each individual changes with time.

\begin{figure}
\centering
\includegraphics[width=3.3 in]{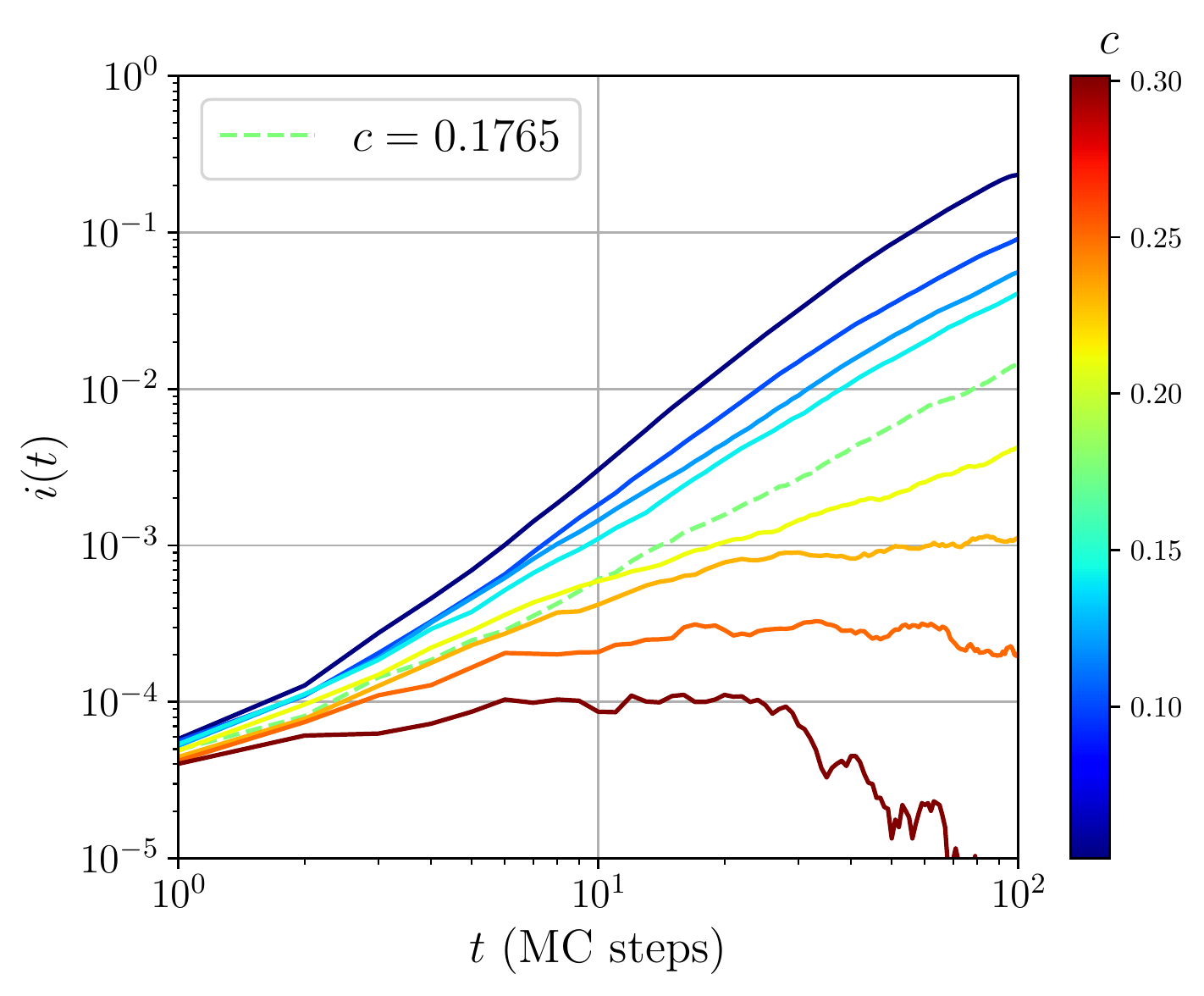}
\caption{Plot of $i(t) \times t$ in $\log \times \log$ scale for different recovery rates $c$ without mobility ($v=0$). The critical point is at $c=0.1765$.}
\label{fig_v0_c1765} 
\end{figure}

In our study, we first identify the phase transition of the model without considering the mobility, i.e., $v=0$. In Fig. \ref{fig_v0_c1765} we show the $\log\times \log$ plot of $i(t) \times t$ for different recovery rates $c$. As presented above, at the phase transition point the behavior of this curve should be linear. 
As expected \cite{Souza2010,Tome2010}, at $c=c_0=0.1765$ (dashed line) we observe such a behavior meaning that this is the critical point for the standard epidemic SIR model. For $c<c_0$ the curves present an almost linear behavior but they last only until $t\approx 10^2$ MC steps. On the other hand, for $c>c_0$ the curves do not present any linear behavior.

\begin{figure}
\includegraphics[width=3.3 in]{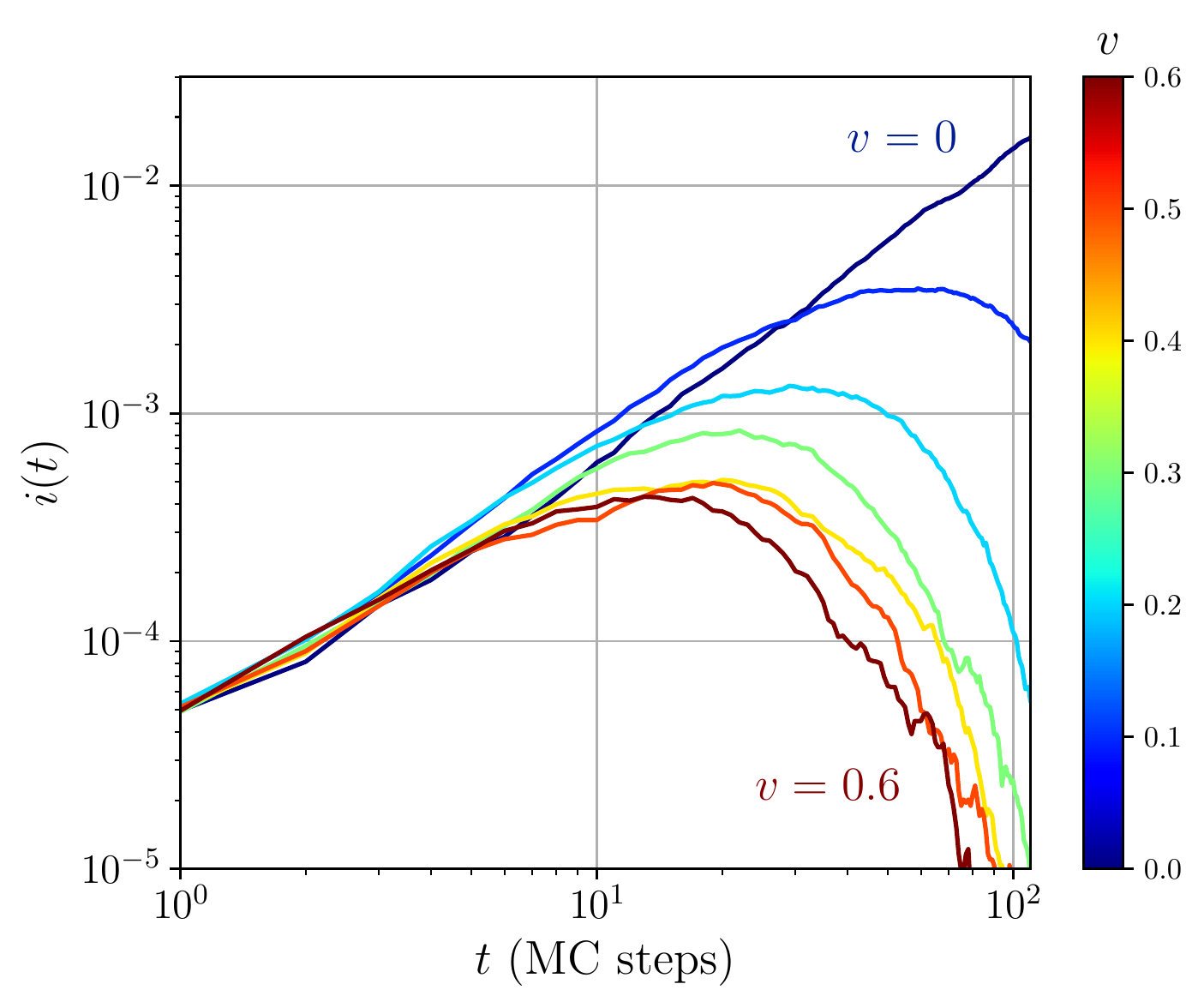}
\caption{The influence of the displacement $v$ on the phase transition point, $c=c_0=0.1765$ of the standard epidemic SIR model. The curves represent the following displacements: $v=0$ (dark blue curve), 0.1, 0.2, 0.3, 0.4, 0.5 and 0.6 (dark red curve).}
\label{fig_c1765_v0_v0p5}
\end{figure}

Figure \ref{fig_c1765_v0_v0p5} shows the behavior of the density of infected individuals as function of time for different mobility $v$ and for $c=c_0$. For $v=0$, the individuals are not able to move and then the standard SIR model is retrieved. In this particular case, the system is at criticality and the curve follows a linear behavior in $\log\times \log$ scale. However, for other values of $v$, even for $v=0.1$ (blue curve), there are no evidence of phase transitions meaning that, at least for this value of $c$, the spreading of the disease occurs only through finite clusters and the outbreak of the epidemic does not occur.
So, for $v>0$, the cluster of infected individuals can grow if the immunization rate $c$ decreases and a possible outbreak of the disease may occur meaning the emergence of a phase transition.

\begin{figure}
\centering
\includegraphics[width=3.3 in]{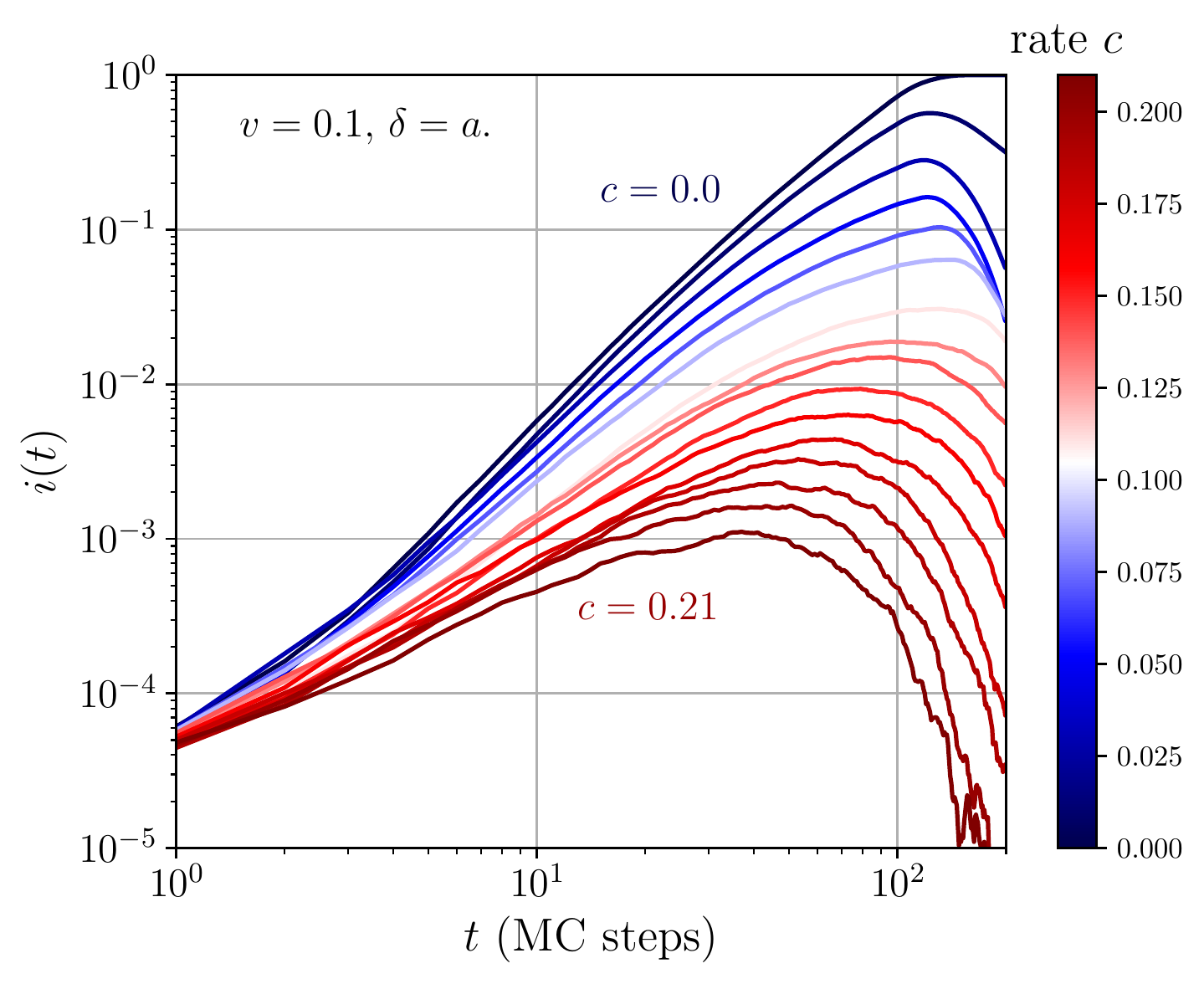}
\caption{The behavior of $i(t) \times t$ in a $\log \times \log$ plot for $v=0.1$ and $\delta = 11a/10$.}
\label{fig_v0p1}
\end{figure}

Now, we fix the value of $v$ at 0.1 and calculate $i(t) \times t$ for different values of the recovery rate $c$ from 0 to 0.21 in order to show how the mobility affects the phase transition. Figure \ref{fig_v0p1} shows that the mobility strongly influences the phase transition and, as in the previous figure, there is no indication of phase transition for any considered recovery rate.

From the results shown above, a natural question arises: For a given $v$, is there any value of the set $c$ and $\delta$ in which the system is at criticality? To answer this question, we carry out nonequilibrium Monte Carlo simulations, as those presented until now, along with the calculation of the coefficient of determination  $\alpha$ in order to find points in which $\alpha(c,\delta) \lesssim 1$, i.e., values of $c$ and $\delta$ which yields a linear behavior in the curves $i(t) \times t$ in $\log \times \log$ scales.

\begin{figure}[!h]
\centering
\subfigure{a)
\includegraphics[width=3.3 in]{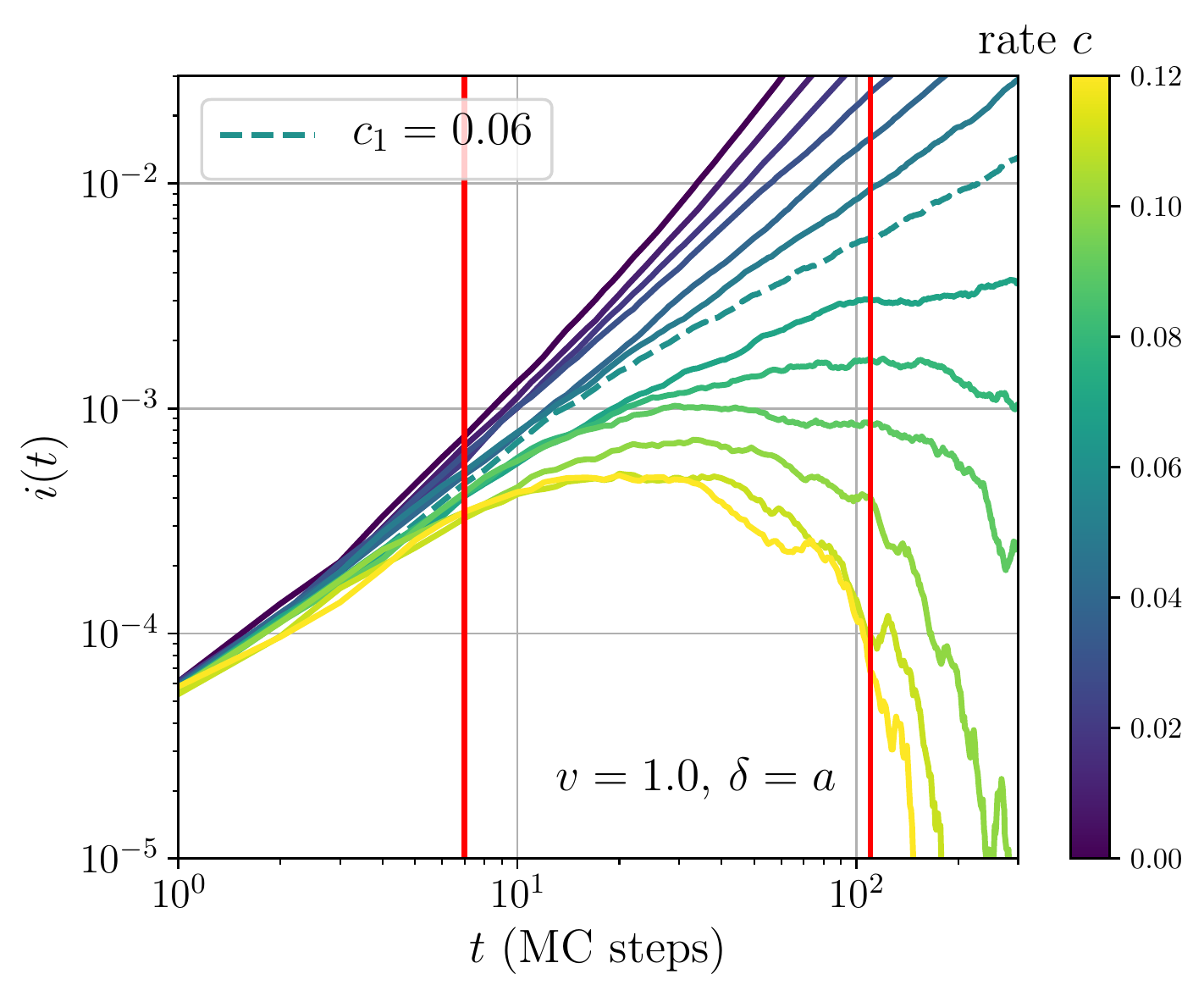}
\label{fig_v1_delta1} 
}
\subfigure{b)
\includegraphics[width=3.3 in]{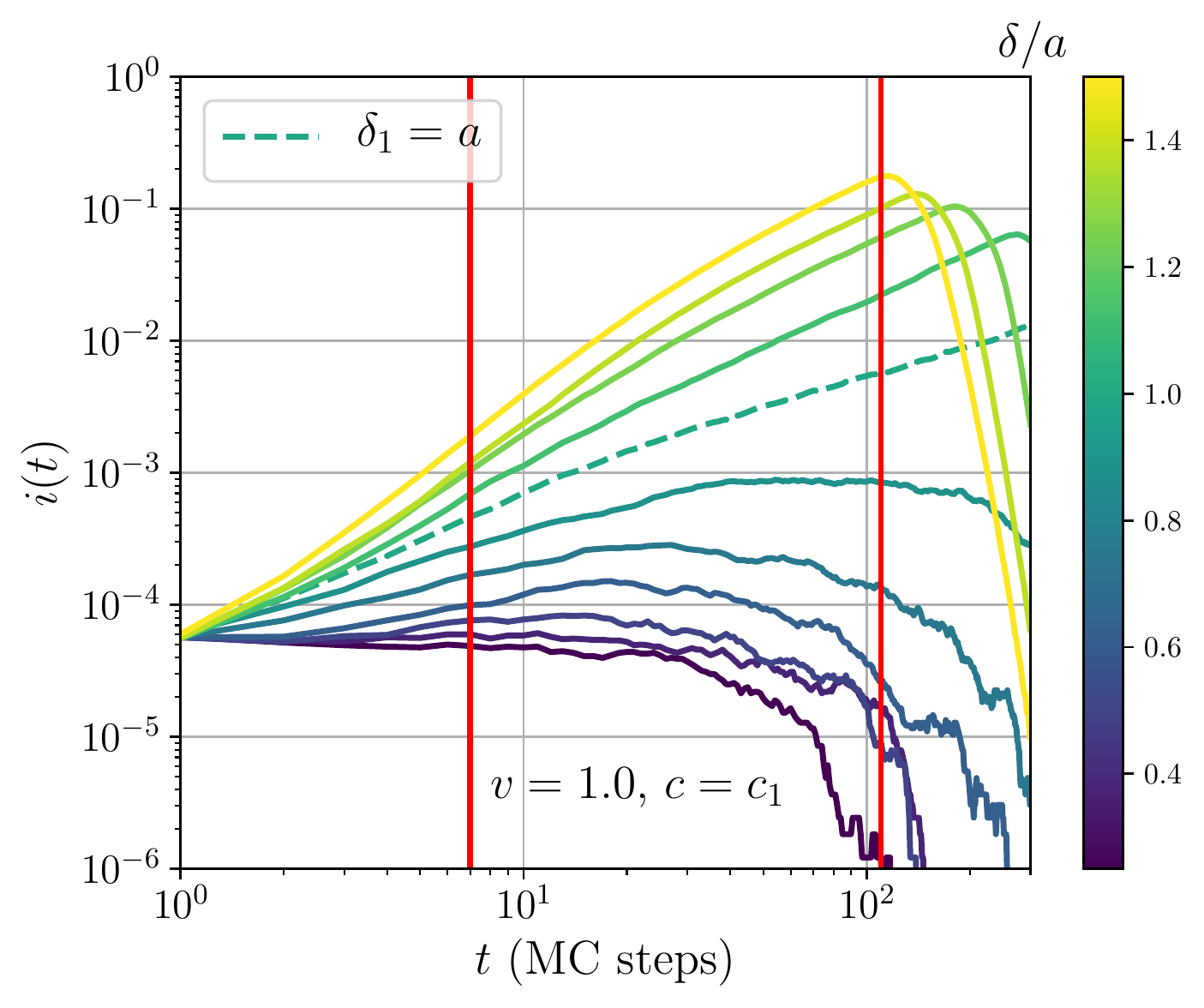}
\label{fig_v1_c0p06}
} 
\caption{Plot of $i(t) \times t$ in log scale around the transition point for $v=1.0$. \subref{fig_v1_delta1} $0<c<0.12$ with step of 0.1 and $\delta = a$. \subref{fig_v1_c0p06} $0.50 < \frac{\delta}{a} < 3.00$ with step of 0.25 and $c_1=0.06$. The red bars correspond to $t_c=7$ and $Q=110$ (see Eq. \ref{coefdeterminacal}).}
\end{figure}

However, before doing this procedure, we present below a further analysis. We look for a critical point by sweeping the recovery rate $c$ from 0 to 0.12 for $v=1.0$ and $\delta = a$. Figure \ref{fig_v1_delta1} shows our results for $i(t) \times t$ in $\log \times \log$ scale. 
As can be seen, there is not a single value of $c$ with linear behavior. Instead, we find straight lines for all curves with $c\geq c_1= 0.06$ (this result is similar to that shown in Fig. \ref{fig_c1765_v0_v0p5}). So, the value $c_1$ is a candidate for the phase transition point. On the other hand, Fig. \ref{fig_v1_c0p06} shows the same plots for different values of $\delta$ with $c = 0.06$ and $v=1.0$. Again, one can see a linear behavior for all curves with $\delta > \delta_1 = a$. So the coordinates $(c_1,\delta_1)$ (within this precision) can be a candidate to a phase transition point, which means that phase transition can exist for $v>0$.

\begin{figure}
\includegraphics[width=3.5 in]{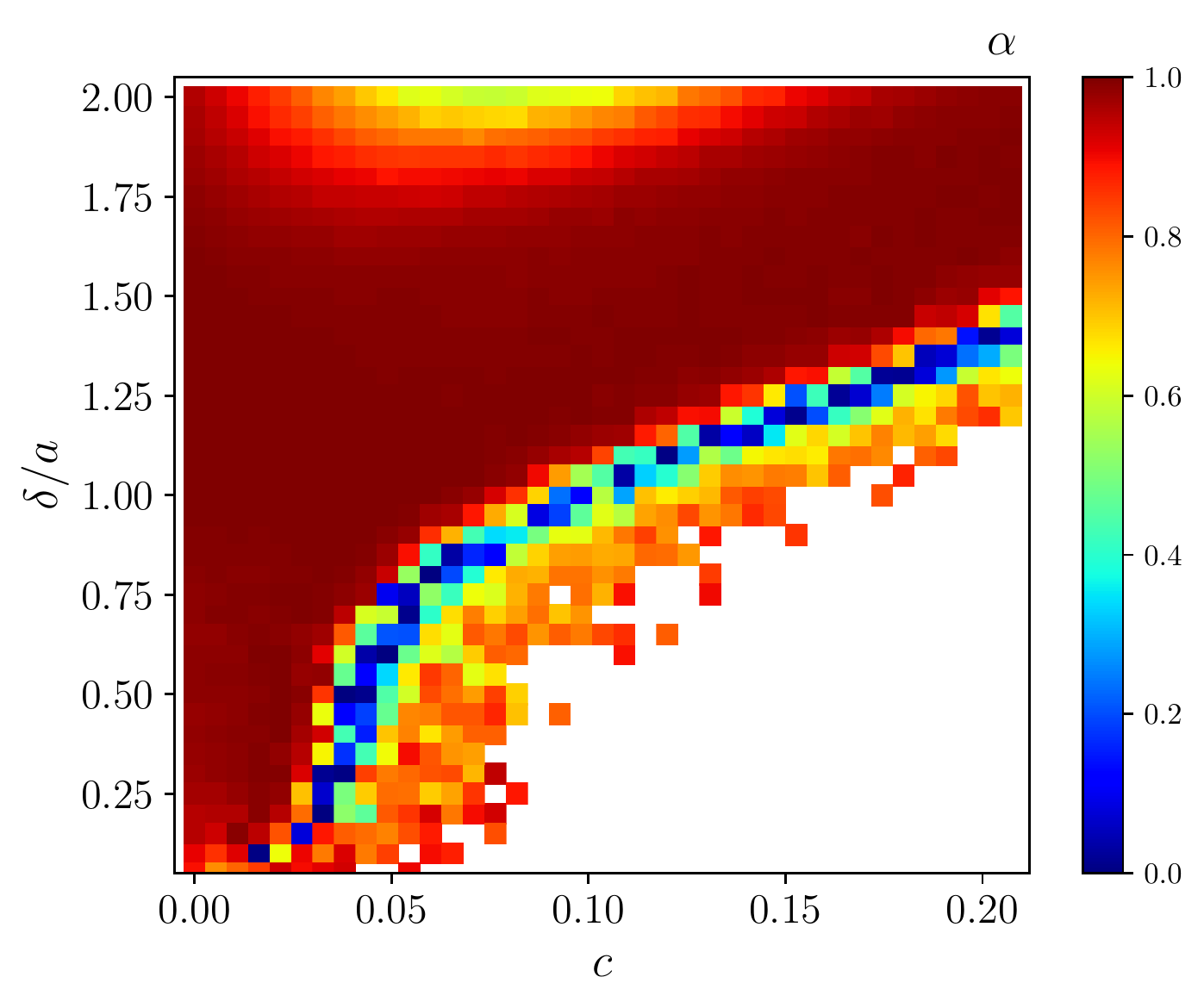}
\caption{Coefficient of determination $\alpha$ (in units of $a$) calculated in the range $0<c<0.2$ and $0.1<\delta/a<2.0$ for $v=1.0$. The red region consists of points where $\alpha \simeq 1$ and the white one refers to points where the linear fit could not be performed since $\ln i(t)$ strongly diverges from a linear behavior.}
\label{fig_alpha_full}
\end{figure}

Now, we finally turn our attention to the coefficient of determination $\alpha$. Figure \ref{fig_alpha_full} shows the color map for $\alpha$ in a large range  of $c$ and $\delta$ with $v=1.0$ ($0<c<0.20$ and $0.1<\delta/a <2.0$). This map consists of 1560 points with steps 0.0055 for $c$ and $0.05$ for $\delta$. The estimate of each $\alpha$ is obtained in the interval $7<t<110$ as indicated by the red vertical bars of Figs. \ref{fig_v1_delta1} and \ref{fig_v1_c0p06}. As shown in Fig. \ref{fig_alpha_full}, there is a large part of the graph with red color, indicating that $\alpha \simeq 1$. Also, there are two borders with other smaller regions where $\alpha < 1$: small yellow-green region on the top and big white area (where the linear fitting was not feasible) at the bottom of this figure. The behavior mimics our findings observed in the previous figures. So, the phase transition points of this extended version of the epidemic SIR model should be right at the borders. 

\begin{figure}[!h]
\centering
\subfigure{a)
\includegraphics[width=3.5 in]{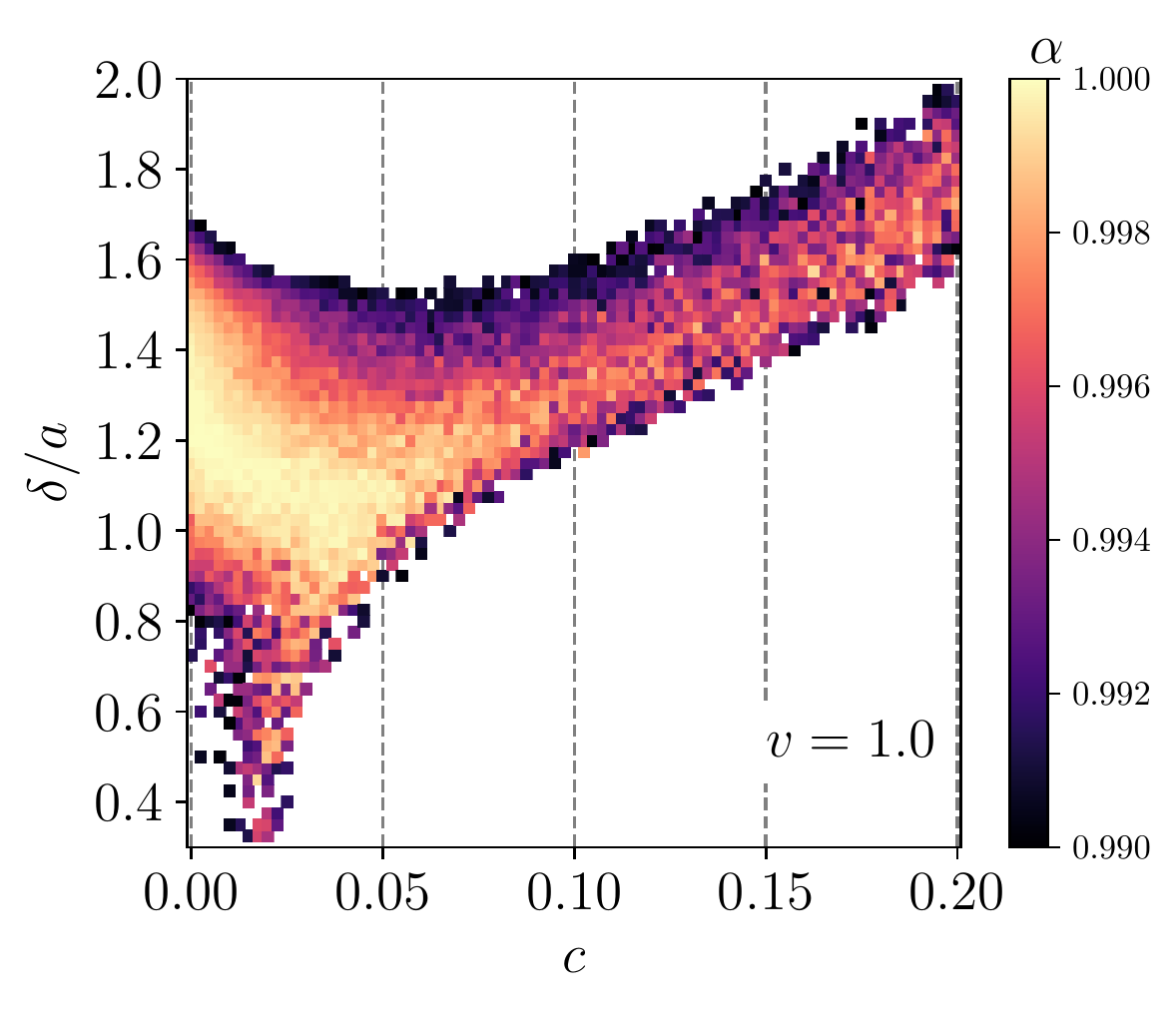}
\label{fig_alpha_13} 
}
\subfigure{b)
\includegraphics[width=3.5 in]{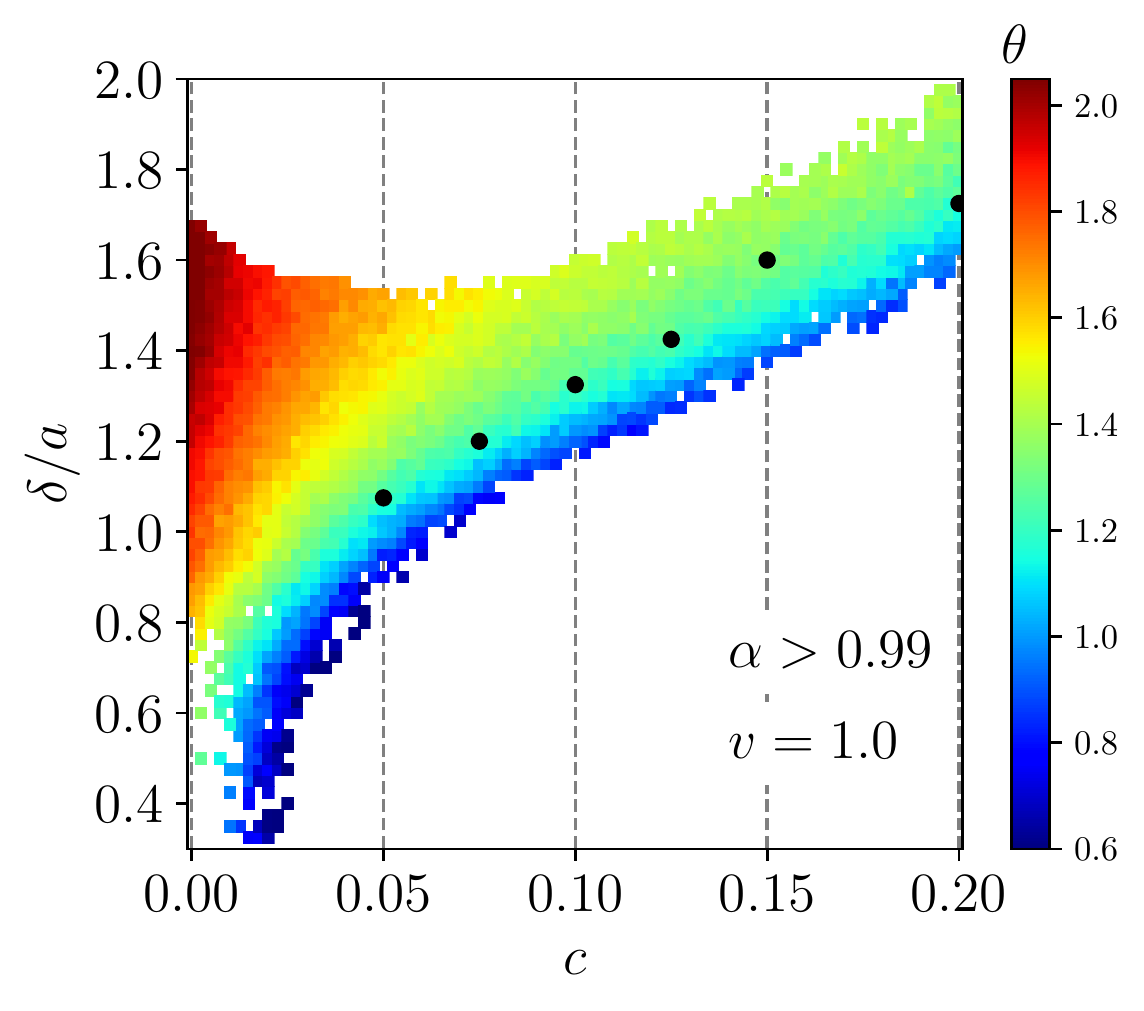}
\label{fig_theta_13}
} 
\caption{ Calculations in the region where $\alpha > 0.99$ for $v=1.0$ with steps 0.0025 for $c$ and 0.025 for $\delta/a$. These figures consist of 2620 points in the selected region for mor accuracy. \subref{fig_alpha_13} Coefficient of determination $\alpha$. \subref{fig_theta_13} Values of the exponent $\theta$. The black points are the points considered in Table \ref{tab:theta}.}
\end{figure}

To interpret Fig. \ref{fig_alpha_full}, we shall consider the candidate phase transition point at $c_1=0.06$ and $\delta_1 = a$, which is right at the lower border. The function $\ln i(t)$ has a linear behavior only for $c \leq c_1$ (dark blue curves in Fig. \ref{fig_v1_delta1}) which is compatible with $\alpha \simeq 1$. In the same way, the linear behavior is only seen with $\delta \geq \delta_1$ according to the Figs. \ref{fig_v1_c0p06} (light yellow lines) and \ref{fig_alpha_full} (red upper part). So, we conclude that, in fact, the phase transition points are at the borders between the regions where $\alpha \simeq 1$ and $\alpha < 1$. This shows that the mobility dependent connectivity can indeed not only remove the phase transition from the value $c_0 = 0.1765$, but it can also create a whole line of critical points. 

To estimate the dynamic critical exponent $\theta$, we look with more detail into the region where the coefficient of determination is close to 1. As shown in Fig. \ref{fig_alpha_13}, there is a large region of points where $\alpha > 0.99$ and which is bounded by points with $\alpha \sim$ 0.99. This figure also shows that inside the region, and along all its extention, there are several points where $\alpha$ is close to one. Figure \ref{fig_theta_13} shows a color map of the exponent $\theta$ for this region. As can be seen, the exponent $\theta$ is varying considerably, i.e., $0.6 \leq \theta \leq 2.0$. Table I presents the exponent $\theta$ and their respective $\alpha$ for seven points ($c$, $\delta/a$) along the critical line (marked as solid black circles in Fig. \ref{fig_theta_13}). This table also shows that the corresponding value of the coefficient of determination for the considered points are all close to one.

\begin{table}[h!]
\centering
\begin{tabular}{ c | c | c |c | c  }
\hline \hline
Point & $c$  & $\delta/a$ & $\alpha  $ & $\theta$  \\
\hline
\hline \#  1 & $0.050$ & $1.075$ & $0.999812543$ & $1.23385131$ \\
\hline \#  2 & $0.075$ & $1.200$ & $0.99878228$ & $1.1756392$ \\
\hline \#  3 & $0.100$ & $1.325$ & $0.997114122$ & $1.25382280$ \\
\hline \#  4 & $0.125$ & $1.425$ & $0.997213125$ & $1.18583548$ \\
\hline \#  5 & $0.150$ & $1.600$ & $0.998348594$ & $1.30938840$ \\
\hline \#  6 & $0.175$ & $ $ & $ $ & $1. $ \\
\hline \#  7 & $0.200$ & $1.725$ & $0.998354495$ & $1.23638952$ \\
\hline
\hline
\end{tabular}
\caption{Values for $\alpha$ and $\theta$ for selected points $(c,\delta)$ on the boundary. Given the coordinate $c$, the coordinate $\delta$ is the one that gives the best value for $\alpha$.}
\label{tab:theta}
\end{table}
As can be seen, the coefficients of determination are very close to 1 for these points. However, the exponents seems to be varying but do not appear to have a pattern of growth, decay or stability. Nevertheless, these results allow us to assert that our model does not belong to the universality class of the original epidemic SIR model with constant connectivity.

\section{Conclusions}
\label{sec:conclusions}

In conclusion, we study the epidemic SIR model with mobility $v$ based on random networks by allowing the individuals (nodes of the network) to perform a random walk. This system considers two features tipically used to describe human interactions: dependency of the number and duration of interactions with time. We showed that, as expected, for $v=0$ our implementation recovers the phase transition point ($c_0=0.1765$) of the original SIR model. However, by considering $v>0$, this transition point was destroyed, raising the following question: Is there a new phase transition point (or region) for a given mobility $v>0$? Tunning the parameters $(c,\delta)$, we successfully found a candidate to the transition point at $(c_1,\delta_1)$ for $v=1.0$. Using the coefficient of determination we found that not only the set $(c_1,\delta_1)$ but a number of them are possible phase transition points. Finally, we obtain different values for the dynamical exponent $\theta$ for some critical points along this curve, which can be an evidence that our model does not follow a universality class. Our results shed light on the mechanisms of the epidemic SIR model when considering features of human interactions on a face-to-face network.

\begin{acknowledgments}

R.S. Grisotto thanks the financial support from the brazilian agency CNPq. The authors gratefully acknowledge the use of the computer resources of the LCC laboratory from Federal University of Goi\'{a}s and of Euler facilities from ICMC/State University of S\~{a}o Paulo. P.F. Gomes also acknowledges the helpful discussions with S. M. Reia.

\end{acknowledgments}

\appendix

\section{Coefficient of determination}
\label{ap:a}

The coefficient of determination for a given set of points $(x_j,y_j)$ with $j=0,1,2,3,...,Q$ is defined as (Eq. (11.12), page 548 of Ref. \cite{Trivedi2002}):
\begin{equation}
\alpha = \dfrac{\sum \left(  \hat{y}_j - \bar{y} \right)^2 }{ \sum \left( y_j - \bar{y} \right)^2 }, \nonumber
\end{equation}
where $\hat{y_j}$ is the predicted value from a given fit and $y_j$ is the observed value (from experiment or simulation). The sum goes from $j=0$ do $j=Q$ and the average is $\bar{y} = (1/(Q+1))\sum y_j$. In our case the points (from simulation) are $(t_j,i_j)$ and the function between them (in the phase transition) is $i(t) = i_0t^{\theta}$. To identify this behavior in the data, we take the natural logarithm to obtain $y = p+ q x$ with:
\begin{equation}
y_j = \ln i_j, \quad x_j = \ln t_j, \quad p = \ln i_0, \quad \textrm{and} \quad q = \theta. \nonumber
\end{equation} 
So, our predicted value is $\hat{y}_j = p+ q x_j$. Here, we want to start the sum in the time $j=j'$ (instead of $j=0$) the average becomes:
\begin{equation}
\ell = \bar{y} = \dfrac{1}{Q-t_{j'}+1}\sum_{j=j'}^Q y_j. \nonumber
\end{equation}  
Using these expressions we obtained the Eq. (\ref{coefdeterminacal}). 

The parameters $q$ (slope) and $p$ were obtained through regular linear regression. Considering a sum from $j=1$ to $j=N$, the expressions are:
\begin{eqnarray}
p &=& \dfrac{ \sum y_j  \sum x_j^2  -  \sum x_j  \sum x_j y_j }{N\sum x_j^2 - \left( \sum x_j \right)^2}, \nonumber \\
q &=& \dfrac{N \sum x_j y_j - \sum x_j \sum y_j  }{N\sum x_j^2 - \left( \sum x_j \right)^2}. \nonumber
\end{eqnarray}

\section{Temporal evolution}
\label{ap:b}

The sequence of the algorithm is described below. A random number is generated many times and is represented by the symbol $\Gamma$.
\begin{enumerate}
\item At $t=0$ the square lattice is generated using the input parameters $N$, $Q$, $v$, $\omega$, $\delta$, $c$, $\rho$, and $N_s$. The values of $L$ and $a$ are calculated. 

\item Each individual is evaluated and if he has infected neighbor(s), the probability of infection $p=(1-c)n_i/n_t$ is calculated. If $\Gamma < p$ and the state of the individual is susceptible, he becomes infected. If he is already infected and $\Gamma < c$, it becomes recovered. 

\item When all individuals are analyzed, the new iteration $t$ begins and the algorithm checks if each individual has other interacting individuals: neighbors within a distance smaller than $\delta$. If it does not, it is displaced by a quantity $\leq v$ in an arbitrary direction $[0,2\pi]$. If it has, it will move only if $\Gamma < \omega$. This is the effect of the interaction: two interacting agents has a chance to remain interacting in their previous vertices. This analysis is performed for all individuals.

\item After a Monte Carlo step, the quantities of interest are estimated and one returns to the step 2.

\end{enumerate}


\begin{thebibliography}{99} 

\bibitem{Fortunato2018} Fortunato S \textit{et al} 2018, \textit{Science} \href{http://dx.doi.org/10.1126/science.aao0185}{\textbf{359} 1007}.


\bibitem{Barabase1999} Barab\'{a}si A L and Albert R 1999 \textit{Science} \href{http://dx.doi.org/10.1126/science.286.5439.509}{\textbf{286} 509}.

\bibitem{Watts1998} Watts D J and Strogatz S H 1998 \textit{Nature} \href{http://dx.doi.org/10.1038/30918}{\textbf{393} 440}.


\bibitem{Wang2014} Wang Z et al 2014 \textit{Wireless Pers Commun} \href{http://dx.doi.org/10.1007/s11277-014-1782-3}{\textbf{78} 755-783}.

\bibitem{Pozzana2017} Pozzana I, Sun K and Perra N 2017 \textit{Phys. Rev. E} \href{https://doi.org/10.1103/PhysRevE.96.042310}{\textbf{96} 042310}.

\bibitem{Starnini2013} Starnini M, Baronchelli A and Pastor-Satorras R 2013 \textit{Phys. Rev. Lett.} \href{http://dx.doi.org/10.1103/PhysRevLett.110.168701}{\textbf{110} 168701}.

\bibitem{Hethcote2000} Hethcote H W 2000 \textit{SIAM Rev.} \textbf{42} 599-653.

\bibitem{PastorSatorras2001} Pastor-Satorras R and Vespignani A 2001 \textit{Phys. Rev. Lett.} \href{http://dx.doi.org/10.1103/PhysRevLett.86.3200}{\textbf{86} 3200}.

\bibitem{Mieghem2013} van Mieghem P and van de Bovenkamp R 2013 \textit{Phys. Rev. Lett.} \href{http://dx.doi.org/10.1103/PhysRevLett.110.108701}{\textbf{110} 108701}.

\bibitem{Earn2000} Earn D J D, Rohani P, Bolker B M, Grenfell B T 2000 \textit{Science} \href{http://dx.doi.org/10.1126/science.287.5453.667}{\textbf{287} 667}.

\bibitem{Vespignani2012} Vespignani A 2012 \textit{Nature Physics} \href{http://www.nature.com/doifinder/10.1038/nphys2160}{\textbf{8} 32}

\bibitem{Newman2002} Newman M E J 2002 \textit{Phys. Rev. E} \href{http://dx.doi.org/10.1103/PhysRevE.66.016128}{\textbf{66} 016128}.

\bibitem{Keeling2005} Keeling M J and Eames K T D 2005 \textit{Jour. of Royal Society Interface} \href{http://dx.doi.org/10.1098/rsif.2005.0051}{\textbf{2} 295–307}.

\bibitem{Genois2015} G\'{e}nois M, Vestergaard C L, Cattuto C and Barrat A 2015 \textit{Nature Communications} \href{http://dx.doi.org/10.1038/ncomms9860}{\textbf{6} 8860}.

\bibitem{Fournet2016} Fournet J and Barrat A 2016 \textit{Scientific Reports} \href{http://dx.doi.org/10.1038/srep24593}{\textbf{6} 24593}.

\bibitem{Silva2015} da Silva R and Fernandes H A 2015 \textit{J. Stat. Mech.} \href{http://dx.doi.org/10.1088/1742-5468/2015/06/P06011}{P06011}.

\bibitem{Barthelemy2004} Barth\'{e}lemy M, Barrat A, Pastor-Satorras R and Vespignani A 2004 \textit{Phys. Rev. Lett.} \href{http://dx.doi.org/10.1103/PhysRevLett.92.178701}{\textbf{92} 178701}. 

\bibitem{Barthelemy2005} Barth\'{e}lemy M, Barrat A, Pastor-Satorras R, Vespignani A 2005 \textit{Journal of Theoretical Biology} \href{http://dx.doi.org/10.1016/j.jtbi.2005.01.011}{\textbf{235} 275–288}.

\bibitem{Hinrichsen2000} Hinrichsen H 2000 \textit{Advances in Physics} \href{http://dx.doi.org/10.1080/00018730050198152}{\textbf{49}:7 815-958}. 

\bibitem{Grassberger1983} Grassberger P 1983 \textit{Math. Biosciences} \textbf{63}: 151-172.

\bibitem{Souza2010} de Souza D R, Tom\'e T 2010 \textit{Physica A} \href{http://dx.doi.org/10.1016/j.physa.2009.10.039}{\textbf{389} 1142-1150}.

\bibitem{Wada2015} Wada A H O, Tom\'e T and de Oliveira M J 2015 \textit{J. Stat. Mech.} \href{http://dx.doi.org/10.1088/1742-5468/2015/04/P04014}{P04014}.

\bibitem{Alexander2015} Alexander H O, Tom\'e T and Oliveira M J 2015 \textit{J. Stat. Mech.} \href{http://dx.doi.org/10.1088/1742-5468/2015/04/P04014}{(2015) P04014}.

\bibitem{Argolo2011} Argolo C, Quintino Y, Gleria I, Lyra M L 2011 \textit{Physica A} \href{http://dx.doi.org/10.1016/j.physa.2010.12.012}{\textbf{390} 1433-1439}.

\bibitem{Colizza2007} Colizza V and Vespignani A 2007 \textit{Phys. Rev. Lett} \href{http://dx.doi.org/10.1103/PhysRevLett.99.148701}{\textbf{99} 148701}.

\bibitem{Estrada2016} Estrada E, Meloni S, Sheerin M, and Moreno Y 2016 \textit{Phys. Rev. E} \href{https://doi.org/10.1103/PhysRevE.94.052316}{\textbf{94} 052316}.

\bibitem{Shu2015} Shu P, Wang W, Tang M and Do Y 2015 \textit{Chaos} \href{http://dx.doi.org/10.1063/1.4922153}{\textbf{25} 063104}.

\bibitem{Newman2010} Newman M E J \textit{Networks: an introduction}, Oxford University Press (2010).

\bibitem{Arashiro2007} Arashiro E and Tom\'e T 2007 \textit{J. Phys. A: Math. Theor.} \href{http://dx.doi.org/10.1088/1751-8113/40/5/002}{\textbf{40} 887-900}.

\bibitem{Juhasz2015} Juh\'{a}sz R, K\'{o}vacs I A and Igloi F 2015 \textit{Phys. Rev. E} \href{http://dx.doi.org/10.1103/PhysRevE.91.032815}{\textbf{91} 032815}.





\bibitem{Belik2011} Belik V, Geisel T and Brockmann D 2011 \textit{Phys. Rev. X} \href{http://dx.doi.org/10.1103/PhysRevX.1.011001}{\textbf{1} 011001}.

\bibitem{Balcan2011} Balcan D and Vespignani A 2011 \textit{Nature Physics} \href{http://www.nature.com/doifinder/10.1038/nphys1944}{\textbf{7} 581-6}.


\bibitem{Ren2014} Ren G and Wang X 2014 \textit{Chaos} \href{http://dx.doi.org/10.1063/1.4876436}{\textbf{24} 023116}.



\bibitem{Alessandretti2017} Alessandretti L, Sun K, Baronchelli A and Perra N 2017 \textit{Phys. Rev. E} \href{https://doi.org/10.1103/PhysRevE.95.052318}{\textbf{95} 052318}.

\bibitem{Scholtes2014} Scholtes I, Wider N, Pfitzner R, Garas A, Tessone C J and Schweitzer 2014 F \textit{Nature Communications} \href{https://doi.org/10.1038/ncomms6024}{\textbf{5} 5024}.

\bibitem{Perra2012} Perra N, Baronchelli A, Mocanu D, Gon\c{c}alves B, Pastor-Satorras R and Vespignani A 2012 \textit{Phys. Rev. Lett.} \href{http://dx.doi.org/10.1103/PhysRevLett.109.238701}{\textbf{109} 238701}.

\bibitem{Dorogovtsev2008} Dorogovtsev S N and Goltsev A V and Mendes J F F 2008 \textit{Rev. Mod. Phys.} \href{http://dx.doi.org/10.1103/RevModPhys.80.1275}{\textbf{80} 1275}.

\bibitem{Tome2010} Tom\'e T and Ziff R M 2010 \textit{Phys. Rev. E} \href{http://dx.doi.org/10.1103/PhysRevE.82.051921}{\textbf{82} 051921 (2010)}.

\bibitem{Trivedi2002} K. S. Trivedi \textit{Probability and Statistics with Realiability}, Queuing, and Computer Science and Applications 2nd edn (New York: Wiley) (2002).

\bibitem{Silva2012} da Silva R, Drugowich de Fel\'{\i}cio J R and Martinez A S 2012 \textit{Phys. Ref. E} \href{https://doi.org/10.1103/PhysRevE.85.066707}{\textbf{85} 066707}.

\bibitem{Silva2013} da Silva R, Fernandes H A, Drugowich de Fel\'{\i}cio J R and Figueiredo W 2013 \textit{Comput. Phys. Commun.} \href{http://dx.doi.org/10.1016/j.cpc.2013.05.005}{\textbf{184} 2371}.

\bibitem{Silva2013b} da Silva R, Alves Jr. N, Drugowich de Fel\'{\i}cio J R 2013 \textit{Phys. Rev. E}  \href{https://doi.org/10.1103/PhysRevE.87.012131}{\textbf{87} 012131}.

\bibitem{Fernandes2014} da Silva R, Fernandes H A and Drugowich de Fel\'{\i}cio F R 2014 \textit{Phys. Rev. E} \href{https://doi.org/10.1103/PhysRevE.90.042101}{\textbf{90} 042101}.

\bibitem{Fernandes2017} Fernandes H A, da Silva R, Drugowich de Fel\'{\i}cio J R and Caparica A A 2017 \textit{Phys. Rev. E} \href{https://doi.org/10.1103/PhysRevE.95.042105}{\textbf{95} 042105}.



\bibitem{Fernandes2016} Fernandes H A, da Silva R, Santos E D, Gomes P F and Arashiro E 2016 \textit{Phys. Rev. E} \href{http://dx.doi.org/10.1103/PhysRevE.94.022129}{\textbf{94} 022129}.


\end{thebibliography}
\end{document}